\documentclass[twocolumn, dvipsnames]{aastex701}

\begin{document}

\title{Revealing Hidden Cosmic Flows through the Zone of Avoidance with Deep Learning}

\author[orcid=0000-0001-9005-2792]{Alexandra Dupuy}
\affiliation{Korea Institute for Advanced Study, 85, Hoegi-ro, Dongdaemun-gu, Seoul, 02455, Republic of Korea}
\email[show]{adupuy@kias.re.kr}

\author[orcid=0000-0002-8434-979X]{Donghui Jeong}
\affiliation{Korea Institute for Advanced Study, 85, Hoegi-ro, Dongdaemun-gu, Seoul, 02455, Republic of Korea}
\affiliation{Department of Astronomy and Astrophysics, and Institute for Gravitation and the Cosmos, The Pennsylvania State University, University Park, PA 16802, USA}
\email{djeong@psu.edu}

\author[orcid=0000-0003-4923-8485]{Sungwook E. Hong}
\affiliation{Korea Astronomy and Space Science Institute, 776 Daedeokdae-ro, Yuseong-gu, Daejeon 34055, Republic of Korea}
\affiliation{Astronomy Campus, University of Science and Technology, 776 Daedeok-daero, Yuseong-gu, Daejeon 34055, Republic of Korea}
\email{swhong@kasi.re.kr}

\author[orcid=0000-0003-3428-7612]{Ho Seong Hwang}
\affiliation{Astronomy Program, Department of Physics and Astronomy, Seoul National University, 1 Gwanak-ro, Gwanak-gu, Seoul 08826, Republic of Korea}
\affiliation{SNU Astronomy Research Center, Seoul National University, 1 Gwanak-ro, Gwanak-gu, Seoul 08826, Republic of Korea}
\affiliation{Australian Astronomical Optics - Macquarie University, 105 Delhi Road, North Ryde, NSW 2113, Australia,}
\email{galaxy79@snu.ac.kr}

\author[orcid=0000-0002-4391-2275]{Juhan Kim}
\affiliation{Center for Advanced Computation, Korea Institute for Advanced Study, 85 Heogiro, Dongdaemun-gu, Seoul, 02455, Republic of Korea}
\email{kjhan@kias.re.kr}

\author[orcid=0000-0003-0509-1776]{Hélène M. Courtois}
\affiliation{Universit\'e Claude Bernard Lyon 1, IUF, IP2I Lyon, 4 rue Enrico Fermi, Villeurbanne, 69622, France}
\email{helene.courtois@univ-lyon1.fr}

\begin{abstract}

We present a refined deep-learning-based method to reconstruct the three-dimensional dark matter density, gravitational potential, and peculiar velocity fields in the Zone of Avoidance (ZOA), a region near the galactic plane with limited observational data.
Using a convolutional neural network (V-Net) trained on A-SIM simulation data, our approach reconstructs density or potential fields from galaxy positions and radial peculiar velocities. The full 3D peculiar velocity field is then derived from the reconstructed potential. We validate the method with mocks that mimic the spatial distribution of the Cosmicflows-4 (CF4) catalog and apply it to actual data. Given CF4's significant observational uncertainties and since our model does not yet account for them, we use peculiar velocities corrected via an existing Hamiltonian Monte Carlo reconstruction, rather than raw catalog distances. 
Our results demonstrate that the reconstructed density field recovers known galaxy clusters detected in an H \textsc{i} survey of the ZOA, despite this dataset not being used in the reconstruction. This agreement underscores the potential of our method to reveal structures in data-sparse regions.
Most notably, streamline convergence and watershed analysis identify a mass concentration consistent with the Great Attractor, at $(l, b) = (308.4^\circ \pm 2.4^\circ, 29.0^\circ \pm 1.9^\circ)$ and $cz = 4960.1 \pm 404.4,{\rm km/s}$, for 64\% of realizations.
Our method is particularly valuable as it does not rely on data point density, enabling accurate reconstruction in data-sparse regions and offering strong potential for future surveys with more extensive galaxy datasets.

\end{abstract}



\section{Introduction}

The large-scale structure of the Universe is shaped by the distribution of dark matter and baryonic matter, forming a cosmic web \citep{1996Natur.380..603B} of walls, filaments, and clusters, and cosmic voids. The motion of galaxies within this web is influenced by gravitational potential wells, leading to peculiar velocities, i.e, deviations from the pure Hubble flow. Peculiar velocities provide a crucial probe of the underlying mass distribution and serve as unbiased dynamical tracers of the total matter in the Universe.

Reconstructing the density and velocity fields from observed peculiar velocities has been a long-standing challenge. Various statistical and Bayesian techniques have been developed to tackle this problem, including the Wiener Filter (WF) and Constrained Realizations (CR) methodologies \citep{1991ApJ...380L...5H, 1999ApJ...520..413Z, 2009LNP...665..565H, 2012ApJ...744...43C, 2024MNRAS.527.3788H}. The WF method offers an optimal linear approach to reconstructing the velocity and density fields and has been instrumental in mapping cosmic structures. It has played a key role in delineating the Laniakea supercluster \citep{2014Natur.513...71T, 2019MNRAS.489L...1D}, or locating the repeller associated with the CMB cold spot \citep{2017ApJ...847L...6C}. More recently, Hamiltonian Monte Carlo (HMC) methods have been introduced as a powerful alternative for reconstructing local density and peculiar velocity fields \citep{2019MNRAS.488.5438G, 2022MNRAS.513.5148V, 2023A&A...670L..15C}. These techniques enabled a more refined cosmographic analysis, allowing the identification of superclusters as gravitational basins \citep{2023A&A...678A.176D, 2024NatAs...8.1610V}.
In addition to methods based on peculiar velocities, a large body of work has focused on reconstructing the density and velocity fields from galaxy redshift surveys \citep{2005ApJ...635...11P, 2006MNRAS.373...45E, 2012MNRAS.427L..35K, 2013MNRAS.432..894J, 2016MNRAS.455.3169L, 2021MNRAS.507.1557L}. These approaches typically rely on Bayesian inference, perturbation theory, or WF to recover the large-scale structure.

Reconstructions from radial peculiar velocities have proven particularly useful for uncovering hidden structures in regions with incomplete data coverage, such as the Zone of Avoidance (ZOA). Such reconstructions have been instrumental in characterizing the Great Attractor \citep[GA, ][]{2013AJ....146...69C} and the Vela supercluster \citep{2019MNRAS.490L..57C}, revealing the gravitational influence of massive structures that were otherwise obscured by the Milky Way.

Recent advances in deep learning have opened new avenues for reconstructing the density and velocity fields. Several studies have employed convolutional neural networks (CNNs) to predict the large-scale structure from galaxy redshift data or reconstructed density fields \citep{2021ApJ...913....2W, 2023JCAP...06..062Q, 2023MNRAS.522.4748W, 2023MNRAS.522.5291G, 2024ApJ...969...76W, 2024A&A...689A.226L}. 
These approaches rely on redshift information rather than directly observed peculiar velocities. In contrast, \cite{2021ApJ...913...76H} introduced a CNN-based method that uses masked galaxy data that contain radial peculiar velocity to reconstruct the density field, demonstrating the potential of deep learning to infer missing information and enhance resolution in sparse datasets.

Building on this idea, we adopt the same CNN architecture to reconstruct fields from radial peculiar velocity data, with two key extensions. First, we apply the model to a significantly larger volume, increasing the sub-volume size from $(40\,{\rm Mpc}/h)^3$ in \citet{2021ApJ...913...76H} to $(160\,{\rm Mpc}/h)^3$. Second, in addition to reconstructing the density field, our network also predicts the gravitational potential, from which we derive the full three-dimensional peculiar velocity field. These extensions allow us to assess the reliability of our method in reconstructing the large-scale structure in the ZOA and to refine the inferred coordinates of the GA, a key region of gravitational influence in the local Universe.

This paper is organized as follows. In Section \ref{sec:data}, we describe the observational and simulation data used in this study, and the generation of training samples from the simulations. Section \ref{sec:methods} details the methodology, including the Deep Learning architecture, training procedure, and evaluation of model performance. Our results are presented in Section \ref{sec:results}, beginning with a discussion on observational uncertainties, followed by reconstructions of the local Universe, an analysis of structures in the ZOA, and an updated localization of the GA. Finally, Section \ref{sec:conclusion} summarizes our findings and outlines potential directions for future work.

\section{Data}
\label{sec:data}

\subsection{Cosmicflows-4}

The Cosmicflows-4 (CF4) catalog \citep{2023ApJ...944...94T} is the most extensive dataset of galaxy distances independent of redshift. CF4 provides precise distance measurements (and consequently radial peculiar velocities) for 55,877 galaxies derived using eight distinct methodologies, with the Tully-Fisher and Fundamental Plane (FP) methods contributing the most substantial data. The catalog offers uniform sky coverage up to 80 Mpc/$h$. Furthermore, two FP samples extend the coverage to greater distances in specific sky regions: the 6dF Galaxy Survey \citep{2014MNRAS.445.2677S}, which extends up to 160 Mpc/$h$ in the southern celestial hemisphere, and the Sloan Digital Sky Survey \citep{2000AJ....120.1579Y, 2022MNRAS.515..953H}, which extends the dataset up to 300 Mpc/$h$. However, coverage is notably sparse in low Galactic latitudes — the ZOA due to obscuration by the Milky Way (MW).

For the rest of the manuscript, we restrict the CF4 sample to galaxies within a cubic volume of width 160 Mpc/$h$ centered on the observer, resulting in a subsample of approximately 17,327 galaxies. 
It is important to note that we do not use the raw observed peculiar velocities from CF4 in our CNN reconstruction, but instead rely on bias-corrected velocities derived from a forward-modeling reconstruction (see Section \ref{sec:HMC} for details).

\begin{figure*}[t!]
\begin{center}
\includegraphics[width=\textwidth,angle=-0]{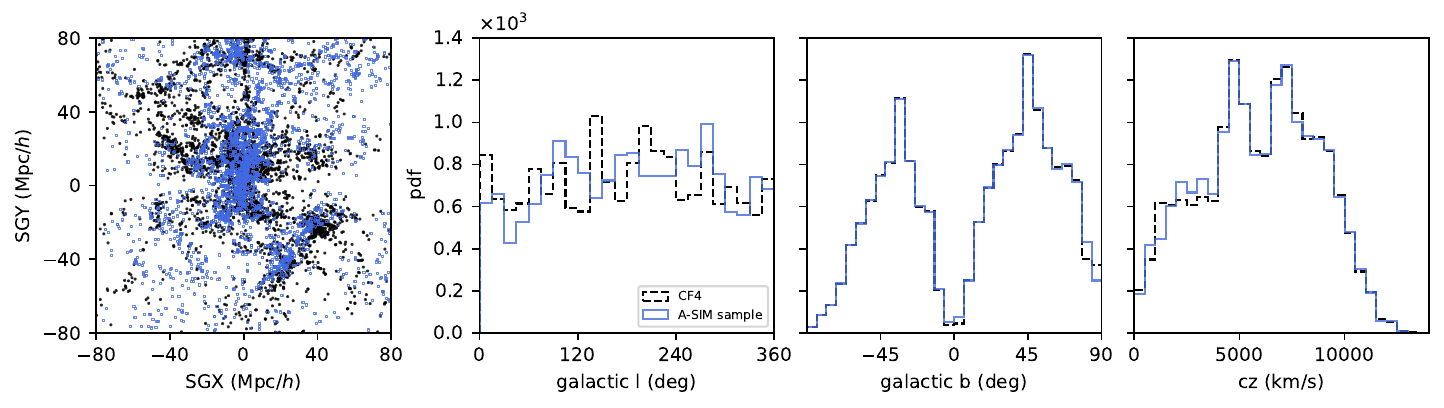}
\caption{Comparison between the CF4 observational data (within the 160 Mpc/$h$ box) and a CF4-like mock sample. From left to right: SGX-SGY slice of width $-15<\mathrm{SGZ}<15$ Mpc/$h$, distributions of galactic longitude $l$, galactic latitude $b$, and redshift $cz$. In each panel, black points and dashed lines represent the CF4 observational data, while blue squares and solid lines correspond to the mock sample.}
\label{fig:mock_CF4}
\end{center}
\end{figure*}

\subsection{A-SIM: $N$-body simulation}

A-SIM is a cosmological $N$-body simulation that was designed to provide a sufficient volume of statistical data for neural network training. The simulation starts at $z=149$ and ends at $z=0$ with 2,980 time steps. The initial conditions are randomly generated with the power spectrum calculated by the CAMB package in the $\mathrm{\Lambda}$CDM cosmological model in a concordance with the WMAP 5-year data \citep{2009ApJS..180..306D}: $(\Omega_\mathrm{m}, \Omega_\mathrm{baryon}, H_0, n_\mathrm{pk}) = (0.26, 0.044, 72 \mathrm{km/s/Mpc}, 0.96)$. In the cubic simulation box of a side length, $L_\textrm{side}=1228.8$ Mpc/$h$, the initial displacement of the particle is calculated according to the second-order Lagrangian Perturbation Theory. Starting from these initial conditions, we simulated the gravitational evolution of the $4096^3$ particles using the Particle-Mesh and Tree methods incorporated into the GOTPM cosmological simulation code \citep{2004NewA....9..111D, 2015JKAS...48..213K}. For the entire simulation run, we spent about 90 days with 1024 CPU cores of the AMD EPYC 7543 32-Core Processor.

From simulation particle data, we built the matter density using the triangular-shape cloud (TSC) method. The gravitational potential on a mesh of $4096^3$ grid cells is calculated by solving the Poisson equation in Fourier space. The galaxy catalog is generated from the merger tree of halos at the 146 time steps based on the most bound particle (MBP)-galaxy abundance matching method \citep{2016ApJ...823..103H} by calculating the merger time scale of satellite halos. 
This method provides the positions and peculiar velocities of galaxies as that of MBPs in halos that are not tidally disrupted, while stellar masses or luminosities are estimated by comparing their corresponding galaxy number density with observations. As a result, A-SIM provides 199,667,555 mock galaxies in the simulation volume at $z = 0$, which corresponds to the galaxy mean number density $\bar{N}_\mathrm{gal} = 1.08 \times 10^{-1} ({\rm Mpc}/h)^{-3}$.

\subsection{Training samples from A-SIM}

The deep learning model considered in this work needs two different input quantities representing, namely, the galaxy positions and their observed peculiar velocities, as well as ``output'' quantities, which is what we want to predict once the model is trained: the dark matter density field and the gravitational potential field. 

Input quantities, namely the galaxy number density $N_\mathrm{gal}$ and the mean peculiar velocity $V_\mathrm{pec}$, are prepared using the A-SIM galaxy catalog. We first identify MW-like galaxies, which serve as the centers of the training samples. 
Since the A-SIM galaxy catalog does not provide stellar masses directly — these can only be assigned through abundance matching with external data — we instead use the TNG100 catalog from the \textit{Illustris-TNG} simulation suite \citep{2018MNRAS.475..676S,2018MNRAS.475..648P,2018MNRAS.475..624N}, where stellar masses are explicitly resolved and thus more reliable for applying a physical mass cut.
Adopting a Milky Way stellar mass of $\sim 5 \times 10^{10}$ M$_\odot$ \citep{2016ARA&A..54..529B,2015ApJ...806...96L}, we select galaxies in TNG100 with stellar masses in the range $4 \times 10^{10}$ M$_\odot < M_* < 1 \times 10^{11}$ M$_\odot$. We then proceed with a resolution correction by matching the cumulative number densities of these galaxies to those in the A-SIM catalog to determine an equivalent stellar mass range. This yields a final selection of MW-like galaxies in A-SIM with stellar masses in the range $4.0 \times 10^{11}$ M$_\odot < M_* < 5.7 \times 10^{11}$ M$_\odot$.

\begin{figure*}[ht!]
\begin{center}
\includegraphics[width=\textwidth,angle=-0]{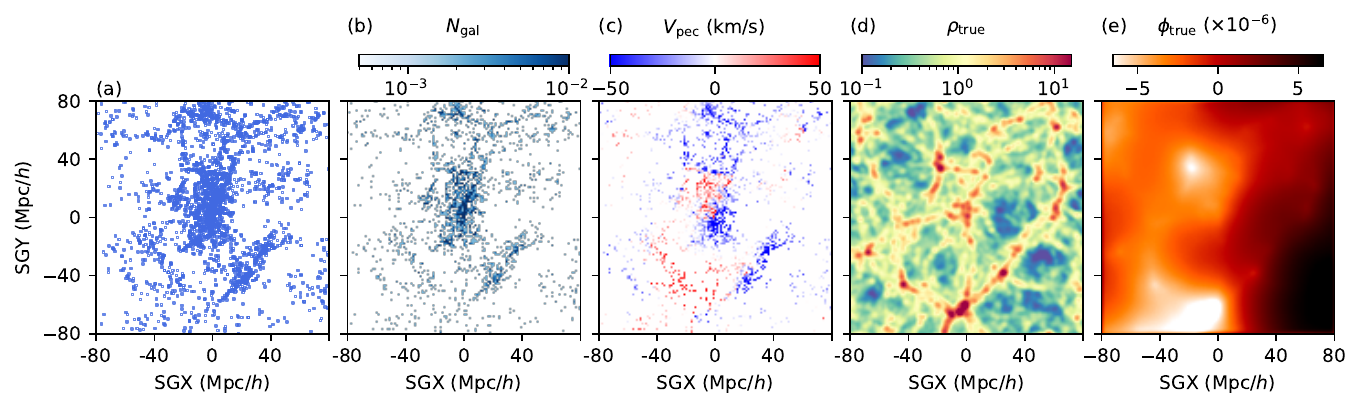}
\caption{Input and output quantities for the Deep Learning algorithm. Each panel corresponds to the same SGX-SGY slice of width $-7.5<\mathrm{SGZ}<7.5$ Mpc/$h$ and of various quantities. From left to right: (a) mock from which the input and output quantities have been generated, (b) galaxy number density $N_\mathrm{gal}$, (c) mean radial peculiar velocity $V_\mathrm{pec}$, (d) true dark matter density field, (e) true gravitational potential field.}
\label{fig:training_sample}
\end{center}
\end{figure*}

To create CF4-like mock samples, each observer is placed at the center of a 160 Mpc/$h$ box. For each CF4 galaxy, we select the closest galaxy from the A-SIM simulation in $(l, b, cz)$ space, using a KDTree. The selection is weighted to give more importance to galactic latitude $b$ and redshift $cz$, ensuring that the mock catalogs reproduce the ZOA and the observed redshift distribution, while galactic longitude $l$ is left essentially free. Once a mock galaxy is selected, it is removed from the candidate pool so that each CF4 galaxy is matched uniquely. This procedure produces mock catalogs with the same number of galaxies as CF4 (17,373), captures the ZOA and redshift distributions, and retains the intrinsic clustering of the simulation subcube rather than enforcing the exact local clustering of the real Universe.
Figure \ref{fig:mock_CF4} presents a detailed comparison between the CF4 observational data (within the 160 Mpc/$h$ box) and a CF4-like mock sample. The figure displays, from left to right, a SGX-SGY slice of width $-15<\mathrm{SGZ}<15$ Mpc/$h$, and the distributions of galactic longitude $l$, galactic latitude $b$, and redshift $cz$. In each panel, black points and dashed lines represent the CF4 observational data, while blue squares and solid lines correspond to the mock sample. The SGX-SGY slice visually represents the spatial distribution of galaxies, showcasing large-scale structures such as clusters, filaments, voids, and the ZOA in both datasets. The distribution of both the black and blue markers shows a clear absence of data points in the ZOA, reflecting how the mock catalog reproduces the lack of coverage in this region. The comparison of the galactic longitude $l$ distributions indicates that the mock catalog effectively captures the variation in galaxy density across different longitudes, although there might be some discrepancies in specific regions due to the difference in large-scale structures. The galactic latitude $b$ distributions show an almost perfect overlap, highlighting that greater weight was given to matching the $b$ distribution rather than $l$ during the construction of the mock. Both $b$ distributions display a noticeable decrease in density at the galactic equator, signifying that the ZOA seen in CF4 is accurately reproduced in the mock data. Finally, the redshift distribution of the mock catalog closely mirrors that of the CF4 data, although there is a slight difference at low redshift due to the low resolution of the simulation, which results in fewer galaxies being represented in the A-SIM galaxy catalog at smaller scales.

Below is described how, for each mock, the input and output quantities for the Deep Learning architecture are generated. Each quantity can be visualized in Figure \ref{fig:training_sample}, where each panel corresponds to the same SGX-SGY slice of width $-7.5<\mathrm{SGZ}<7.5$ Mpc/$h$ of various quantities. The mock from which the input and output quantities shown in this Figure have been generated can be visualized in panel (a), where each blue marker corresponds to a galaxy.

From the CF4-like mock samples, two \(128^3\) cubes, each with a side length of 160 Mpc/$h$, are generated. The galaxies are positioned at their Supergalactic Cartesian coordinates (SGX, SGY, SGZ) computed from $V_\mathrm{cmb}$. The galaxy number density cube $N_\mathrm{gal}$ is then constructed by counting the number of galaxies located in each voxel. Similarly, the mean peculiar velocity cube $V_\mathrm{pec}$ is constructed by averaging in each voxel the line-of-sight peculiar velocity of galaxies, derived from the three-dimensional peculiar velocity provided by the A-SIM galaxy catalog. Both $N_\mathrm{gal}$ and $V_\mathrm{pec}$ can be visualized in panels (b) and (c) in Figure \ref{fig:training_sample}, respectively.

To construct the output quantities, namely the dark matter density field and the gravitational potential field, $128^3$ cubes with a side length of 160 Mpc/$h$ are extracted from the A-SIM density and potential fields, respectively, centering each cube on a MW-like galaxy identified from the galaxy catalog. The original resolution of the simulation grids being 0.3 Mpc/$h$, all cubes have to be resampled to the desired resolution of 1.25 Mpc/$h$, done by first applying Gaussian smoothing with a 1.25 Mpc/$h$ radius to the original higher-resolution grid and subsequently downsampling to obtain \(128^3\) cubes of side length 160 Mpc/$h$. Both output quantities are displayed in panels (d) and (e) of Figure \ref{fig:training_sample}. 

We use in total 11,512 samples to train the deep learning model, divided into 10,248 samples for the training set and 1,264 samples for the validation set.

\subsection{Note on observational uncertainties}
\label{sec:HMC}

\begin{figure}[t!]
\begin{center}
\includegraphics[width=0.9\columnwidth,angle=-0]{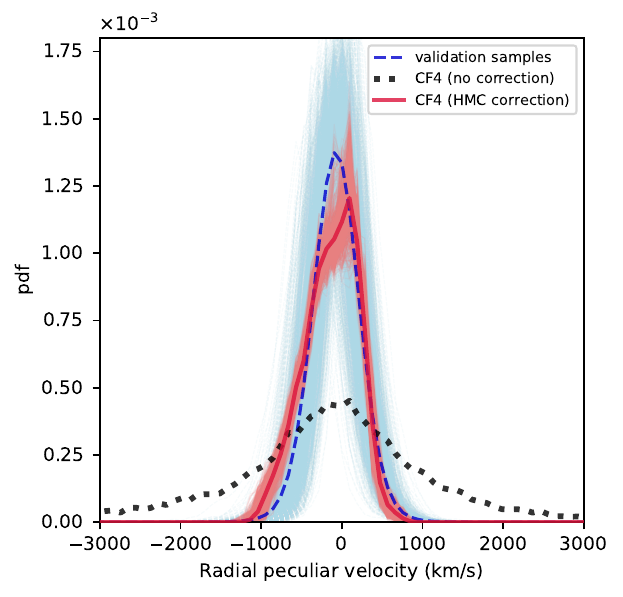}
\caption{Distribution of radial peculiar velocities. The blue dashed lines represent the A-SIM validation samples, with thin lines showing individual samples and the thicker line indicating the mean distribution. The red solid lines correspond to the CF4 dataset after correction using HMC reconstruction, with thin lines representing individual samples and the thicker line showing the mean. The black dotted line depicts the CF4 distribution before correction.}
\label{fig:CF4_vpec_HMC}
\end{center}
\end{figure}

Peculiar velocity catalogs, such as CF4, play a crucial role in understanding the large-scale structure of the Universe. However, working with observed peculiar velocity data presents significant challenges due to inherent limitations and biases. These catalogs are often characterized by large observational uncertainties and systematic errors, which can substantially impact the accuracy of the inferred velocity fields. A key limitation is Malmquist bias, which arises from distance-dependent selection effects, leading to overrepresentation of intrinsically brighter or more easily detectable objects. Additionally, cosmic variance affects the representativeness of local peculiar velocity measurements, while zero-point calibration errors can introduce systematic deviations. Anisotropies in the data coverage and the presence of measurement errors in distance indicators further compound these challenges.

To illustrate these challenges, Figure \ref{fig:CF4_vpec_HMC} shows the distribution of radial peculiar velocities derived from CF4 distances (black dotted line). These velocities are computed using the relation $v_\mathrm{p} = cz - H_0 d$, where $d$ is the galaxy distance from the CF4 catalog, $cz$ is the velocity in the Cosmic Microwave Background (CMB) frame ($V_\mathrm{cmb}$), and $H_0$ is the Hubble constant, set to $H_0 = 75$ km/s/Mpc (best-fitting $H_0$ value for CF4 according to \cite{2023ApJ...944...94T}). 

We compare this distribution to that of the validation samples (blue dashed lines), which represent an idealized case without observational uncertainties and serve as a reference for the expected radial peculiar velocity input distribution for our CNN model. Thin lines correspond to individual validation samples, and the thicker line indicates their mean distribution. In contrast to these idealized inputs, the CF4-derived radial peculiar velocities exhibit a significantly broader distribution, primarily due to observational uncertainties, including distance measurement errors.

Correcting the CF4 data by addressing biases is particularly challenging due to its compilation from many different datasets, each with distinct characteristics. There is no clear selection function for the combined catalog, making it difficult to account for observational biases uniformly across the data. Furthermore, galaxy distance measurements in CF4 are derived using eight different methodologies, each with varying levels of uncertainty and systematic errors. These differences in measurement techniques introduce inconsistencies that complicate efforts to harmonize the data and treat biases effectively. Given these complexities, directly applying the observed CF4 peculiar velocities to the deep learning model is not advisable yet, as the model, in its current state, does not incorporate mechanisms to account for observational uncertainties.

The peculiar velocities used in our analysis are extracted from an existing velocity field reconstructed from CF4 data \citep{Courtois2023}. This reconstruction is based on a Hamiltonian Monte Carlo (HMC) methodology \citep{2019MNRAS.488.5438G}, which incorporates observational CF4 data along with several prior assumptions. The velocity field is derived using a WF approach but is computed iteratively. Specifically, a Bayesian parameter space is defined, including quantities such as galaxy distance and the velocity field itself. At each Markov Chain Monte Carlo (MCMC) iteration, these parameters are updated, and a realization of the velocity field is computed accordingly. Since galaxy positions, distances, and peculiar velocities are iteratively refined based on current estimates, this procedure effectively corrects for uncertainties on observed distances. The final output consists of approximately 10,000 realizations of the velocity and overdensity fields. 

\begin{figure*}[ht!]
\begin{center}
\includegraphics[width=\textwidth,angle=-0]{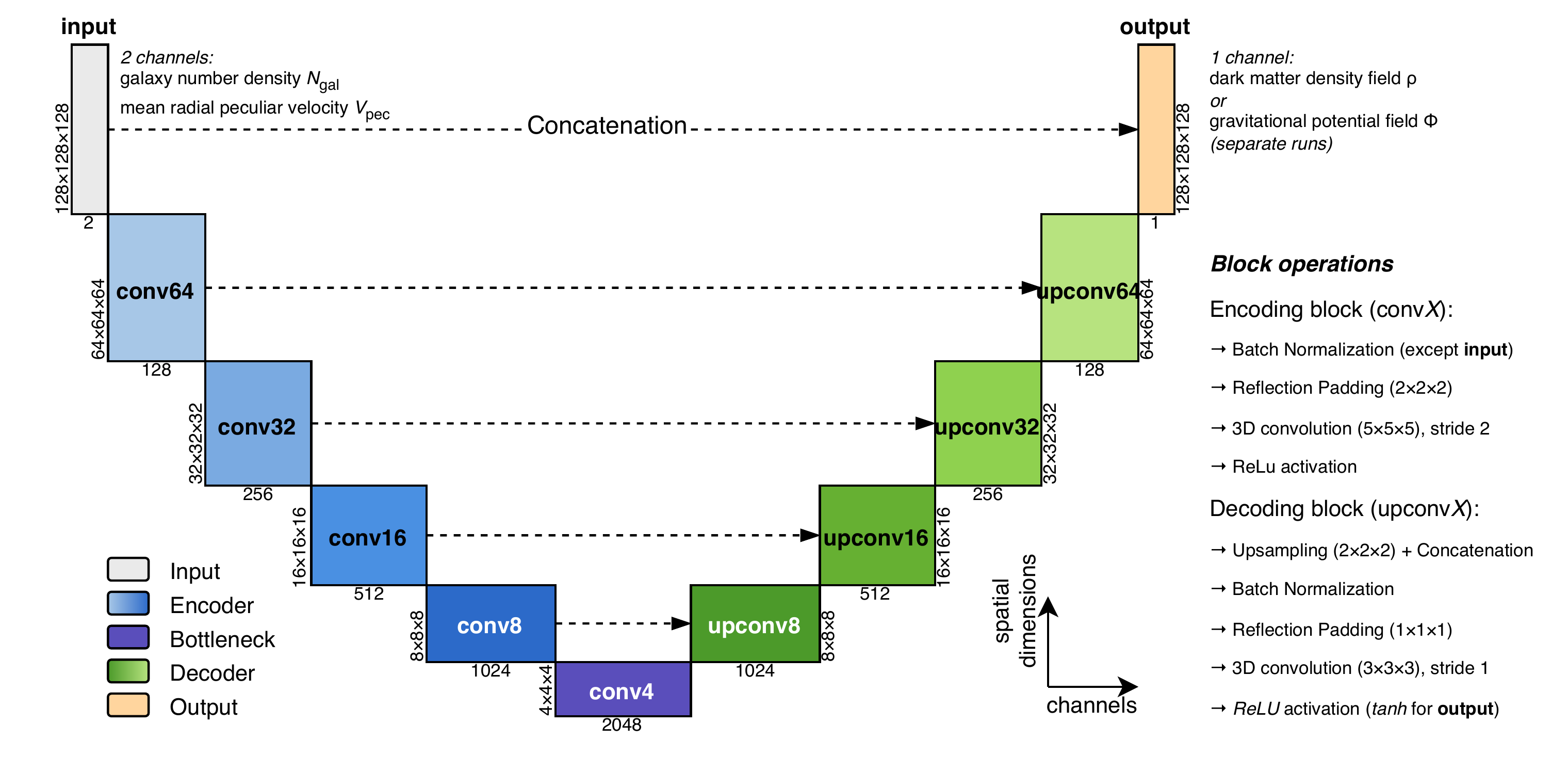}
\caption{Architecture of the 3D V-Net used in this work. The input has two channels: galaxy number density $N_{\mathrm{gal}}$ and mean radial peculiar velocity $V_{\mathrm{pec}}$. Spatial features are encoded via convolutional layers (\texttt{convX}), followed by a bottleneck layer and a symmetric decoder path (\texttt{upconvX}) with skip connections. The output is a single channel representing either the dark matter density field $\rho$ or the gravitational potential field $\phi$, trained separately using the same input and architecture. The width and height of each block indicate the number of channels (below) and spatial dimensions (left), respectively. The detailed sub-operations within each encoding and decoding block are summarized on the right.}
\label{fig:unet_architecture}
\end{center}
\end{figure*}

From the full set of 10,000 HMC velocity field realizations, we randomly select 1,000. For each selected realization, we extract the three-dimensional peculiar velocity values at the positions of the CF4 galaxies and project them along the line of sight to obtain radial peculiar velocities. This process yields one corrected CF4 sample per realization, resulting in a total of 1,000 corrected CF4 samples. The distribution of corrected radial peculiar velocities is shown in Figure \ref{fig:CF4_vpec_HMC} as red solid lines, with thin lines representing individual samples and the thicker line showing the mean. This corrected distribution is significantly narrower than the original CF4 dataset, closely matching the training samples used in our CNN model.

\section{Methods}
\label{sec:methods}

\subsection{CNN architecture and training process}

Our approach employs a deep learning model based on a 3D V-Net CNN, designed for processing volumetric data, to infer the local density and gravitational potential fields from peculiar velocity data. This architecture was originally introduced by \cite{2021ApJ...913...76H} for density field reconstruction; we adopt here a similar model with key enhancements. Specifically, we extend the method by introducing separate training runs for the density and gravitational potential fields — the latter being a novel addition. Furthermore, we increase the size of the reconstructed volume from $(40\,{\rm Mpc}/h)^3$ in the original work to $(160\,{\rm Mpc}/h)^3$, while maintaining the same grid size of $128^3$, resulting in a lower spatial resolution of $1.25\,{\rm Mpc}/h$.
The model follows an encoder-decoder structure, where the encoding and decoding phases progressively transform the input data through a series of convolutional and upsampling layers. The input tensor has a shape of $(2, 128, 128, 128)$, representing a 3D volume with two channels: the galaxy number density $N_\mathrm{gal}$ and the mean radial peculiar velocity $V_\mathrm{pec}$, as described in the previous section. A schematic overview of the full network architecture is shown in Figure \ref{fig:unet_architecture}.

The encoding phase sequentially reduces the spatial dimensions of the input while increasing the depth, extracting progressively large-scale patterns through a series of convolutional layers. Each encoding step follows a structured process: batch normalization is applied (except for the 2-channel input layer) to stabilize training, followed by 2-pixel reflection padding in all three dimensions to preserve spatial structure. A 3D convolution with a $5 \times 5 \times 5$ kernel and a stride of 2 then reduces the spatial dimensions by half while increasing the number of channels. This hierarchical encoding process is repeated across five convolutional layers, progressively reducing the spatial size of the input while expanding the feature depth from 128 channels to 2048 in the final encoding layer.

Once the network has extracted essential spatial patterns from the input, the decoding phase reverses this transformation, reconstructing the original spatial dimensions while progressively reducing the number of channels. Each decoding step begins with upsampling, which doubles the spatial dimensions of the input. To recover small-scale structures, the upsampled features are concatenated with the corresponding output from the encoding phase. Batch normalization is then applied to the concatenated features, followed by 1-pixel reflection padding to maintain spatial consistency. A 3D convolution with a $3 \times 3 \times 3$ kernel and a stride of 1 is then used, followed by a \texttt{ReLU} activation function (except in the final output layer). The decoding phase consists of four such layers, gradually refining the features while decreasing the number of channels from 1024 to 128.

The final layer completes the reconstruction by upsampling and concatenating with the original input, before applying a 3D convolution with a \texttt{tanh} activation function. This produces the output tensor of shape $(1, 128, 128, 128)$, with values normalized between $-1$ and $+1$. By leveraging reflection padding and skip connections through concatenation layers, the architecture effectively captures spatial hierarchies and preserves small-scale information, making it well-suited for reconstructing complex 3D data.

Given the computational demands of this deep learning model, we train it separately for each target field. Specifically, we conduct one training process where the model predicts the dark matter density $\rho$ and another independent training where the output is the gravitational potential field $\phi$. This separation ensures optimized learning for each physical quantity while maintaining model efficiency.

The loss function to be minimized during training is defined as the Mean Square Error (MSE) between true $y_X^\mathrm{true}$ and predicted $y_X^\mathrm{pred}$ fields, where the subscript $X$ can denote either the dark matter density field $\rho$ or the gravitational potential field $\phi$. The MSE is then defined such as:
\begin{equation}
    \mathcal{L}_\mathrm{MSE} = \frac{1}{n} \sum_{i=1}^{n} \left( y_X^{i,\mathrm{pred}} - y_X^{i,\mathrm{true}} \right)^2,
\end{equation}
where the sum goes through the $n$ samples the loss is computed on. 

A simple normalization is applied to both output fields $\rho$ and $\phi$ in order to obtain values ranging between $-1$ and $+1$:
\begin{equation}
    y_\rho = \frac{1}{2.5} \log_{10} \left( \frac{\rho}{\bar{\rho}} \right)
\end{equation}
for the dark matter density field and:
\begin{equation}
    y_\phi = \left( \phi - \bar{\phi} \right) \times \left( 8 \times 10^4 \right)
\end{equation}
for the gravitational potential field. Quantities $\bar{\rho}$ and $\bar{\phi}$ denote the mean density and potential, respectively.

\begin{figure}[ht!]
\begin{center}
\includegraphics[width=0.9\columnwidth,angle=-0]{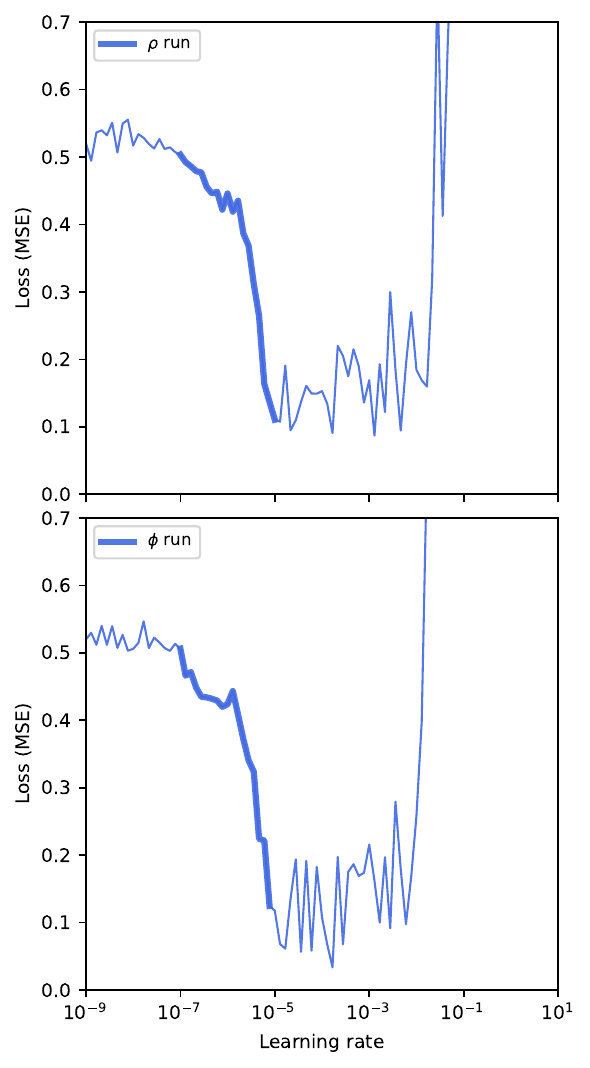}
\caption{Learning rate range test showing the evolution of the loss $\mathcal{L}_\mathrm{MSE}$ as a function of the learning rate $\alpha$. Top: test training on the dark matter density field. Bottom: test training on the gravitational potential field. The adequate range for the triangular cyclic learning rate used for the final training is shown as a thicker solid line on each panel.}
\label{fig:LR_test}
\end{center}
\end{figure}

Instead of using a fixed value for the learning rate, we apply a triangular cyclic learning rate scheme in order to avoid the training to be stuck in a local minima:
\begin{equation}
    \alpha_t = \alpha_\mathrm{L} + \frac{\alpha_\mathrm{U} - \alpha_\mathrm{L}}{T/2} \times \mathrm{min} \left\{ \left( t \; \mathrm{mod} \; T \right), T - \left( t \; \mathrm{mod} \; T \right) \right\},
\end{equation}
where $T = 8$ is the number of mini-batches considered during a single learning rate cycle. The possible range of values for the learning rate is delimited by the minimum $\alpha_\mathrm{L}$ and maximum $\alpha_\mathrm{U}$ learning rate values. A suitable range of learning rates is found by performing a learning rate range test, which consists of few quick training tests of 5 epochs each for various fixed learning rate values. 

Results of the learning rate range test for each output, $\rho$ and $\phi$, can be seen in the top and bottom panels of Figure \ref{fig:LR_test}, respectively. The value of the loss $\mathcal{L}_\mathrm{MSE}$ at the final epoch (5th epoch) is plotted as a function of the value of the learning rate $\alpha$ considered for each test training. For low learning rates, i.e, the parameter vector update is too slow, the loss function as a function of the learning rate $\mathcal{L}_\mathrm{MSE}(\alpha)$ has a flat slope, meaning that the CNN is unable to learn. Inversely, for high learning rate values, the interval of the parameter vector update is too large to find a solution, showing an exploding loss. A suitable $\left[ \alpha_\mathrm{L}, \alpha_\mathrm{U}\right]$ range would be from a learning rate value $\alpha_\mathrm{L}$ large enough for the CNN to start learning, i.e no more flat slope for $\mathcal{L}_\mathrm{MSE}(\alpha)$, to a value $\alpha_\mathrm{U}$ small enough, before the noisy increment of the loss at large $\alpha$. The range for the triangular cyclic learning rate is then fixed as such:
\begin{itemize}
    \item $\alpha_\mathrm{L}^\rho = 10^{-7}$ and $\alpha_\mathrm{U}^\rho = 10^{-5}$ for the training on the dark matter density field $\rho$,
    \item $\alpha_\mathrm{L}^\phi = 10^{-7}$ and $\alpha_\mathrm{U}^\phi = 7 \times 10^{-6}$ for the training on the gravitational potential field $\phi$,
\end{itemize}
and can be visualized as a thicker solid line in Figure \ref{fig:LR_test}.

Finally, the CNN deep learning model is trained using a minibatch size of 8 over 200 epochs, with each epoch consisting of 157 minibatches. The training process utilizes the \texttt{ADAM} optimizer to adjust the model parameters. To ensure optimal performance and mitigate overfitting, model checkpoints are saved at minimum training loss and minimum validation loss. Additionally, the model is also saved at the final epoch. The training process is conducted on two \texttt{NVIDIA A100} GPUs, each equipped with 80GB of HBM2e memory, providing substantial computational power and memory capacity to handle the intensive 3D convolutional operations.

\begin{figure}[ht!]
\begin{center}
\includegraphics[width=0.9\columnwidth,angle=-0]{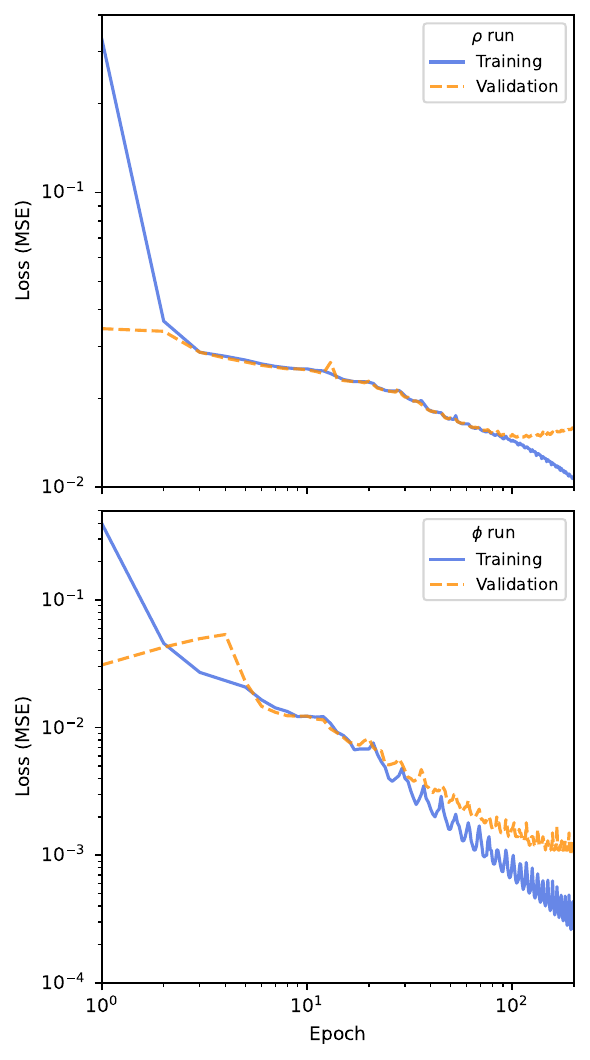}
\caption{Evolution of the loss $\mathcal{L}_\mathrm{MSE}$ as a function of the epoch for the training on the dark matter density field $\rho$ (top) and the gravitational potential field $\phi$ (bottom). The training loss is shown as a blue solid line, while the validation loss is represented by an orange dashed line.}
\label{fig:MSE_epoch}
\end{center}
\end{figure}

Figure \ref{fig:MSE_epoch} reports the evolution of the evolution of the loss $\mathcal{L}_\mathrm{MSE}$ as a function of the epoch, during the training on the dark matter density field $\rho$ (top) and the gravitational potential field $\phi$ (bottom) over 200 epochs each. The training loss is shown as a blue solid line while the validation loss is shown as an orange dashed line. 

For the training on $\rho$, both losses initially start high and decrease rapidly in the first few epochs. The training loss continues to decrease steadily, while the validation loss follows a similar trend but begins to increase slightly towards later epochs. This divergence suggests potential overfitting, where the model performs better on the training data than on the validation data. However, the validation loss does not increase significantly towards the later epochs, indicating only a slight divergence from the training loss. This suggests that the model does not overfit substantially. The close alignment of training and validation losses throughout most of the epochs implies that the model generalizes well to unseen data. Therefore, the results are unlikely to suffer significantly from overfitting, and the model maintains good performance on both the training and validation datasets.

\begin{figure*}[th!]
\begin{center}
\includegraphics[width=0.9\textwidth,angle=-0]{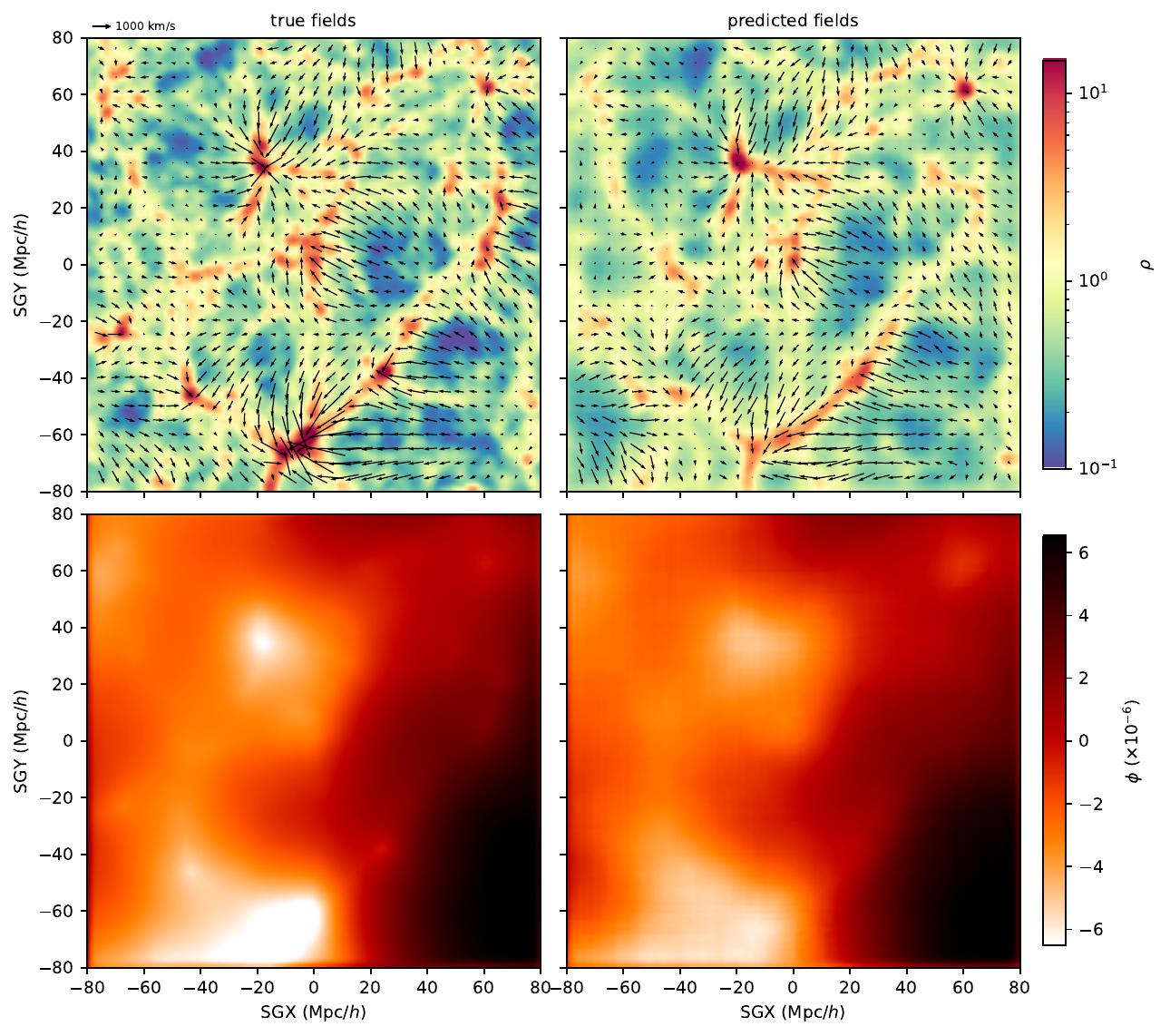}
\caption{Comparison of the dark matter density (top row) and gravitational potential (bottom row) fields. Left panels show the true fields, and right panels show the fields predicted by the CNN model. Predictions are obtained from a single, randomly chosen validation sample. All panels display the same SGX-SGY slice of width $-15<\mathrm{SGZ}<15$ Mpc/$h$. The true and predicted velocity field, derived from each respective potential field, is shown on top of each density field (arrows).}
\label{fig:all_models}
\end{center}
\end{figure*}

As for $\phi$, both losses start high and decrease rapidly within the first few epochs. The training loss continues to decrease steadily, while the validation loss seems to be stabilizing at the last few epochs. The validation loss remains higher than the training loss towards the later epochs, but does not exhibit a substantial increase, indicating minimal overfitting. Notably, both the training and validation loss curves are significantly noisy, showing frequent oscillations. This noise likely stems from the choice of the triangular cyclic learning rate range $[\alpha_\mathrm{L}^\phi, \alpha_\mathrm{U}^\phi]$. The cyclic variation of the learning rate can lead to fluctuations in the loss values as the model experiences phases of rapid learning and slower convergence. While this approach can help the model escape local minima and potentially improve generalization, it also introduces higher variance in the loss curves, as seen in this panel.

\subsection{Performance of the CNN model}

\begin{figure*}[ht!]
\begin{center}
\includegraphics[width=\textwidth,angle=-0]{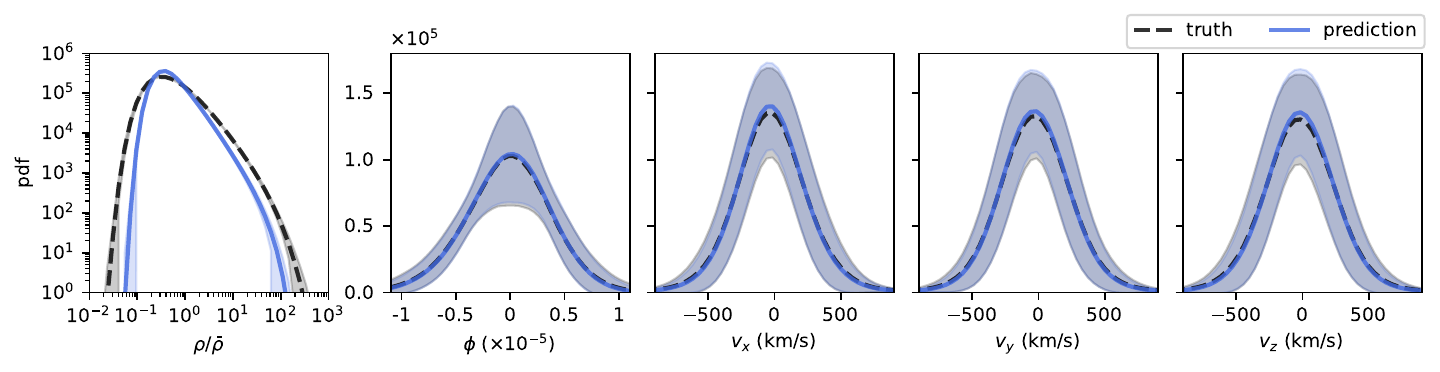}
\caption{Probability density functions (PDFs) of the true (black dashed lines and transparent bands) and predicted (blue solid lines and transparent bands) fields, from left to right: density field, potential field, cartesian components of the 3D peculiar velocity field. The lines and bands represent, respectively, the mean and $1\sigma$ standard deviation over all validation samples.}
\label{fig:histograms}
\end{center}
\end{figure*}

\begin{figure}[h!]
\begin{center}
\includegraphics[width=0.84\columnwidth,angle=-0]{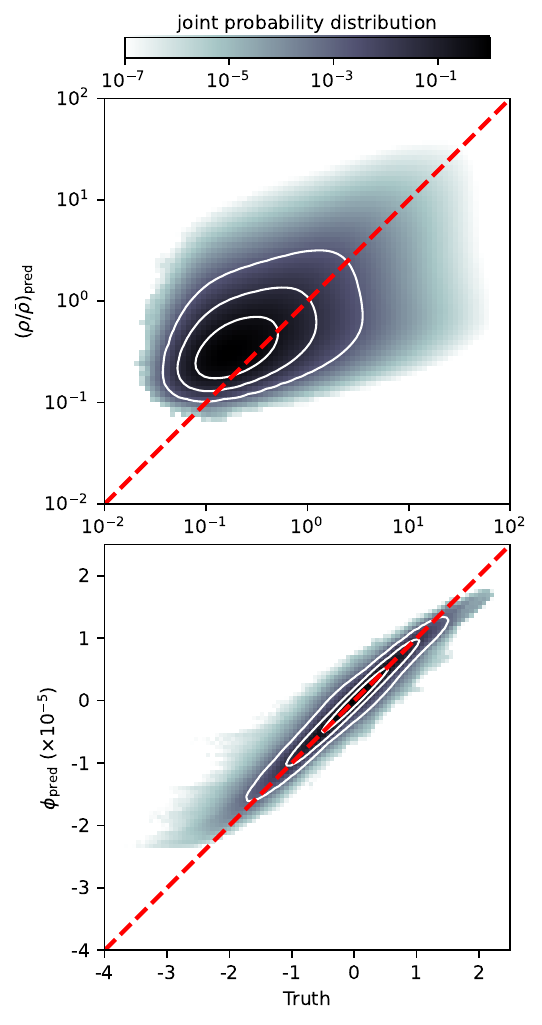}
\caption{Joint probability distribution of the predicted and true values with 1, 2, and 3$\sigma$ certainty level contours (white solid lines), derived from all validation samples, for both density (top), and potential (bottom) fields. The identity line is shown on each panel as a dashed red line for reference. Pearson correlation coefficients between predicted and true fields: $0.65 \pm 0.03$ for density, $0.98 \pm 0.01$ for potential, and although not shown in the Figure, $0.83 \pm 0.12$ for the $x$, $0.81 \pm 0.14$ for the $y$ and $0.83 \pm 0.12$ for the $z$ components of the velocity field (mean and $1\sigma$ standard deviation across all validation samples).}
\label{fig:joint_pdf}
\end{center}
\end{figure}

\begin{figure*}[th!]
\begin{center}
\includegraphics[width=0.9\textwidth,angle=-0]{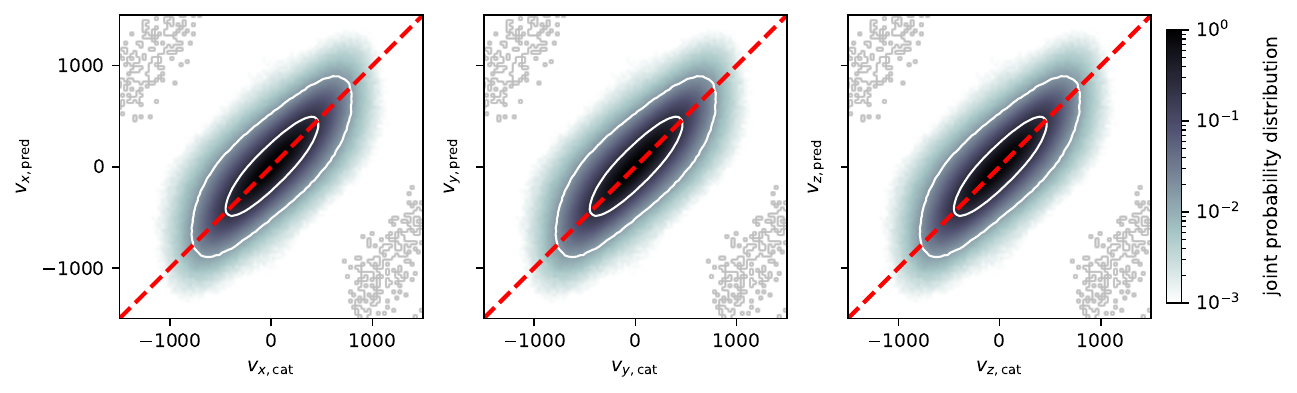}
\caption{Joint probability distribution of the predicted peculiar velocity extracted at the positions of galaxies in all validation samples and the peculiar velocity values listed in the A-SIM galaxy catalog. The three panels correspond to the $x$, $y$, and $z$ components of the peculiar velocity from left to right, respectively. The white solid lines represent the $1\sigma$ and $2\sigma$ confidence level contours of the distribution. The red dashed diagonal line represents the one-to-one relation for reference. Pearson correlation coefficients between predicted peculiar velocities and those from the galaxy catalog: $0.52 \pm 0.11$ for the $x$, $0.43 \pm 0.12$ for the $y$ and $0.48 \pm 0.12$ for the $z$ components of the velocity field (mean and $1\sigma$ standard deviation across all validation samples).}
\label{fig:joint_pdf_v} 
\end{center}
\end{figure*}

We now turn to the performance of the trained CNN model on the validation samples, before applying the model to the CF4 data. The model was saved at the final epoch (epoch 200 for both the $\rho$ and $\phi$ models) and at two additional checkpoints: one at the epoch with the minimum training loss (epoch 200 for $\rho$ and epoch 193 for $\phi$) and one at the epoch with the minimum validation loss (epoch 113 for $\rho$ and epoch 193 for $\phi$). For simplicity, and since all checkpoints yield very similar results, we only consider the model at the final epoch in this manuscript.

Figure \ref{fig:all_models} shows a visual comparison between the true fields (left) and the fields predicted by the CNN model at the final epoch (right). Predictions are obtained by applying the CNN models (for $\rho$ and $\phi$) to a single, randomly chosen validation sample. All panels display the same SGX-SGY slice of width $-7.5<\mathrm{SGZ}<7.5$ Mpc/$h$. The top row compares the true dark matter density field $\rho_\mathrm{true}$ with the predicted field $\rho_\mathrm{pred}$. The true and predicted fields are visually very similar, with the CNN model successfully reconstructing the overdensities (in red) and underdensities (in blue) present in the original density field, even within the ZOA (SGY $\sim$ 0 Mpc/$h$). However, we observe that the predicted large-scale structures are smoother, and the overdensities appear less pronounced compared to the true field. The bottom row compares the true gravitational potential field $\phi_\mathrm{true}$ with the predicted field $\phi_\mathrm{pred}$. As with the density field, the CNN model accurately reconstructs the potential wells (blue) and hills (red). A smoothing effect is also visible in the predicted potential field.

At redshift $z=0$, from the gravitational potential field, the three-dimensional peculiar velocity field $\vec{v}$ can be derived through: 
\begin{equation}
    \vec{v} = - \frac{2 f}{3 H_0 \Omega_m} \vec{\nabla}{\phi},
\end{equation}
where $H_0 = 72$ km/s/Mpc is the Hubble constant, $\Omega_m = 0.26$ is the matter density parameter and $f = \Omega_m^\gamma$ is the growth rate of large-scale structures, where $\gamma=0.55$ in the $\Lambda$CDM model.  

The peculiar velocity field derived from each (true and predicted) potential field is shown in the top row of Figure \ref{fig:all_models}, represented by arrows overlaid on the corresponding dark matter density field. For both the true and predicted fields, the velocity field aligns with its respective potential field, with arrows pointing toward potential wells and away from potential hills. The velocity flows observed in the true (derived) velocity field are also well reconstructed in the predicted velocity field. Notably, the velocity field derived from the predicted potential field matches the predicted density field, even though they are predicted separately, demonstrating that the CNN model successfully captures the relationship between the two fields despite the separate training runs. 

Figure \ref{fig:histograms} compares the probability density functions (PDFs) of the true (black dashed lines and transparent bands) and predicted (blue solid lines and transparent bands) fields: dark matter density field, gravitational potential field, and cartesian components $x$, $y$ and $z$ of the three-dimensional velocity field, from left to right, respectively. The lines represent the mean over all 1,264 validation samples, while the transparent bands around each curve represent the $1\sigma$ standard deviation. The predicted fields closely follow the associated truths, with slight deviations as shown by the overlapping transparent bands.

As an additional performance test, Figure \ref{fig:joint_pdf} presents the joint probability distribution of the predicted and true values for both the density field (top), and the potential field (bottom), across all 1,264 validation samples. The color scale represents the density of points in logarithmic space, with darker regions indicating a higher probability density. Overlaid on the distribution are white contour lines marking the 1$\sigma$, 2$\sigma$, and 3$\sigma$ confidence intervals. The red dashed identity line (truth $=$ prediction) serves as a reference for perfect agreement. While the predictions for the potential field closely align with the identity line, indicating strong agreement, the density field exhibits a broader scatter, especially in high-density regions, which is consistent with the deviations observed in Figure \ref{fig:histograms}.

Figure \ref{fig:joint_pdf_v} presents the joint probability distributions of the predicted peculiar velocity extracted at galaxy positions across all 1,264 validation samples, compared against the peculiar velocity values listed in the A-SIM galaxy catalog. This differs from Figure \ref{fig:joint_pdf}, where the joint probability distributions of the density and potential fields were computed across all grid voxels. The three panels correspond to the $x$, $y$, and $z$ components of the peculiar velocity from left to right, respectively. The white solid contours represent the $1\sigma$ and $2\sigma$ confidence levels. The red dashed diagonal line represents the ideal one-to-one correspondence for reference. The overall structure of the distributions suggests a strong agreement between predicted peculiar velocities and those listed in the A-SIM galaxy catalog. 

To quantify the linear agreement between predictions and true values, we compute the Pearson correlation coefficient, which ranges from $-1$ (perfect anti-correlation) to $1$ (perfect correlation), with $0$ indicating no linear correlation. As reported in the figure captions, the coefficients confirm the trends observed visually: the potential field predictions exhibit an almost perfect linear relationship with the true field, while the density field shows a moderately strong correlation, consistent with the scatter observed in Figure \ref{fig:joint_pdf}. The density field is more nonlinear and contains stronger small-scale variations compared to the smoother gravitational potential, making it inherently more challenging for the network to reconstruct accurately. The coefficients for the velocity components, although not shown visually in the Figure, are included for completeness and indicate strong correlations. 

When comparing predicted velocities to the ones listed in the A-SIM galaxy catalog, the coefficients are moderate, reflecting the increased difficulty of reconstructing the galaxy velocities, which include additional local motions beyond the large-scale gravitational flow. For reference, when comparing the predictions to the gridded true velocities extracted at the same galaxy positions, the coefficients are substantially higher ($0.79 \pm 0.08$, $0.80 \pm 0.10$ and $0.79 \pm 0.10$ for the $x$, $y$ and $z$ components, respectively), indicating that the network captures the underlying large-scale velocity field more accurately. These results are consistent with the expectation that galaxy peculiar velocities are inherently noisier due to small-scale, local dynamics, whereas the smoothed gridded velocities predominantly trace the coherent gravitational flow.

From a given three-dimensional peculiar velocity field, one can easily compute the bulk flow, which is defined by the volume-weighted average of the peculiar velocity field inside a sphere of a given radius $R$ centered on the observer:
\begin{equation}
    \vec{V}_\mathrm{bulk} (R) = \frac{3}{4\pi R^3} \int_{r<R} \vec{v}(\vec{r})d\vec{r}.
\label{eq:bulkflow}
\end{equation}

\begin{figure}[h!]
\begin{center}
\includegraphics[width=0.85\columnwidth,angle=-0]{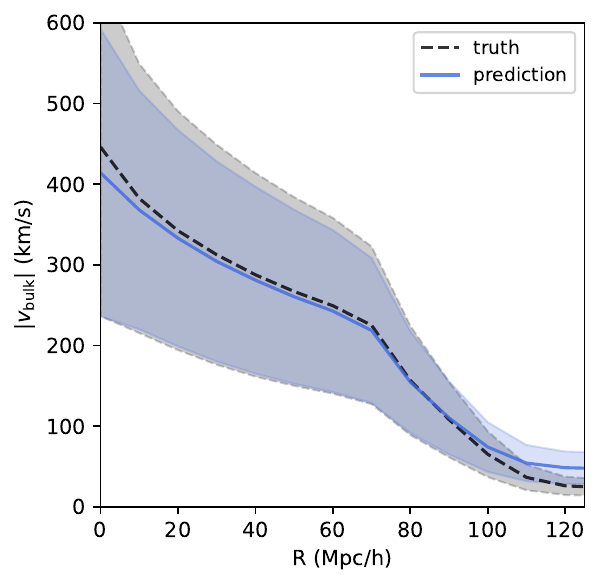}
\caption{Norm of the bulk flow as a function of the sphere radius $R$. The black dashed line represents the mean bulk flow computed from the true velocity fields across all validation samples, while the blue solid line corresponds to the mean bulk flow derived from the predicted velocity fields. The transparent bands indicate the standard deviation.}
\label{fig:vbulk}
\end{center}
\end{figure}

Figure \ref{fig:vbulk} illustrates the bulk flow amplitude as a function of scale, comparing the true bulk flow with that recovered from the predicted velocity fields. The predicted bulk flow closely follows the true bulk flow, demonstrating the accuracy of the reconstructions. At large scales, the bulk flow amplitude decreases and approaches zero, consistent with the cosmological principle, as peculiar velocities average out over large volumes. The transparent bands indicate the standard deviation, which represents cosmic variance: larger at small scales due to the local cosmography and velocity fluctuations, and decreasing at larger scales as the velocity field becomes more isotropic. On average, the predicted bulk flow deviates from the true value by approximately 23 km/s, value estimated across all validation samples and radius bins, demonstrating robust recovery of the true bulk flow across scales.

\section{Results}
\label{sec:results}

\begin{figure*}[hp!]
\begin{center}
\includegraphics[width=\textwidth,angle=-0]{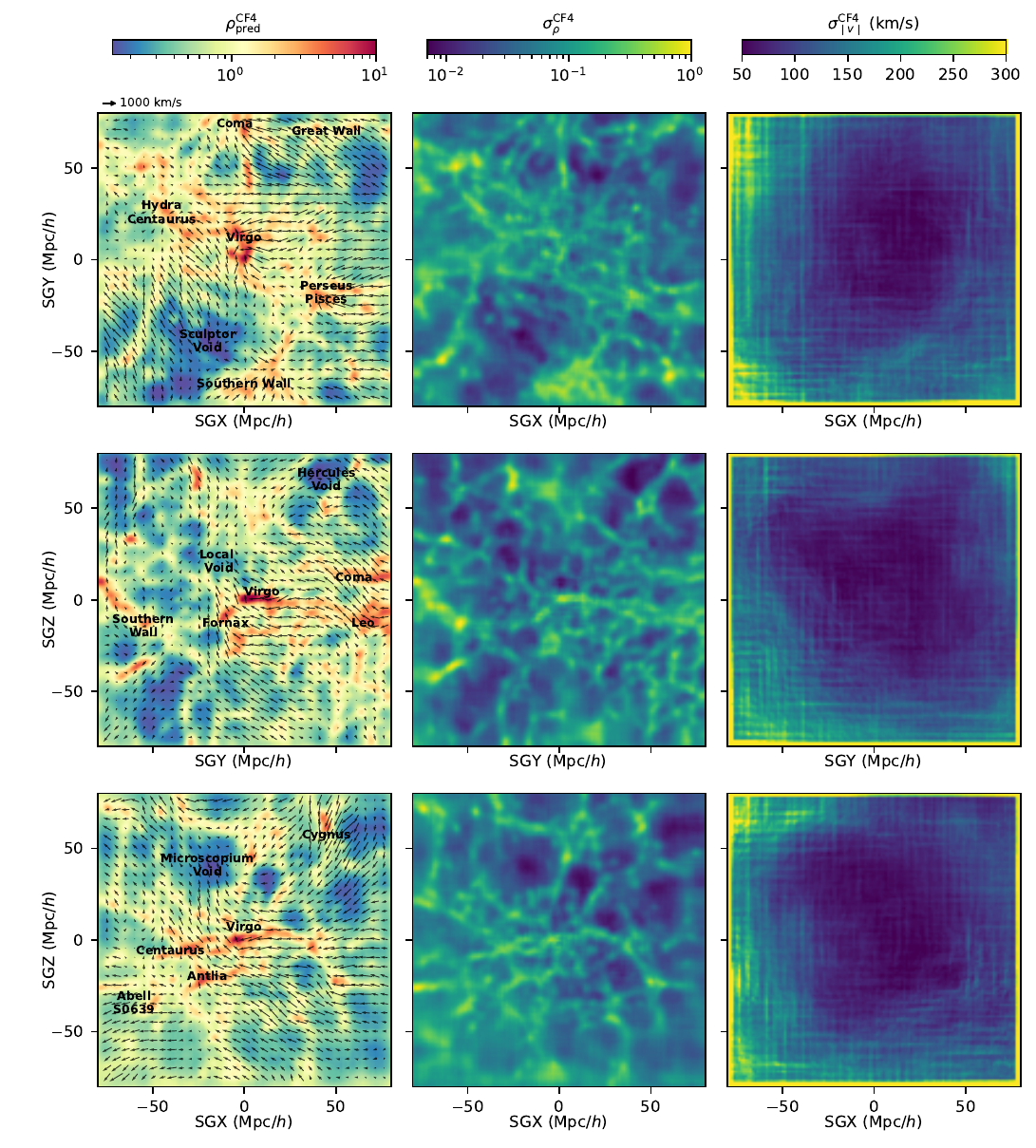}
\caption{Reconstructed large-scale structures in Supergalactic coordinates from 1,000 CNN predictions on CF4. Left column: mean density (colormap: blue for underdensities, red for overdensities) and peculiar velocity (arrows) fields on slices of width $15\,{\rm Mpc}/h$: SGX-SGY and SGY-SGZ (top and middle panel, centered on the observer), and SGX-SGZ (bottom panel, shifted to SGY = $15\,{\rm Mpc}/h$). Middle and right columns: standard deviation of the density and velocity fields, respectively, for the same three slices, with higher uncertainties in yellow.}
\label{fig:CF4cosmography} 
\end{center}
\end{figure*}

\subsection{Local Universe from Cosmicflows-4}

To assess the large-scale structure reconstructed from the CF4 data, we apply our CNN model to the ensemble of 1,000 corrected CF4 samples, producing 1,000 corresponding predictions of the local density and gravitational potential fields. This statistical approach accounts for observational uncertainties and enables a robust estimation of cosmic structures.

The left column of Figure \ref{fig:CF4cosmography} presents slices of the reconstructed large-scale structures in Supergalactic coordinates, showing the mean density and velocity fields across 1,000 realizations predicted by our CNN model, applied for CF4 corrected samples. The top panel displays an SGX-SGY slice with a width of $15\,{\rm Mpc}/h$, centered on SGZ = $0\,{\rm Mpc}/h$, while the middle panel shows an SGY-SGZ slice of the same width, centered on SGX = $0\,{\rm Mpc}/h$. The bottom panel features an SGX-SGZ slice, also $15\,{\rm Mpc}/h$ wide, but shifted to SGY = $15\,{\rm Mpc}/h$ to highlight structures above the ZOA in +SGY. The colormap represents the mean reconstructed density field, with overdensities (clusters) shown in red and underdensities (voids) in blue. Major cosmic structures are labeled according to their associated density features. Overlaid arrows indicate the mean reconstructed peculiar velocity field across all realizations, demonstrating strong alignment with the reconstructed densities despite being predicted independently.

The middle and right columns of Figure \ref{fig:CF4cosmography} display standard deviation maps that quantify the uncertainties across all 1,000 realizations, using a color scale from dark blue (low uncertainty) to yellow (high uncertainty), for the same three Supergalactic slices as the left column. The middle column shows the standard deviation of the density field, with higher values around overdense regions, where variations between realizations are more pronounced. The right column presents the standard deviation of the velocity field norm, which exhibits a different uncertainty pattern, with larger values in regions farther from the observer where data constraints are weaker. Overall, the standard deviations remain within reasonable ranges, indicating the robustness of the reconstructed fields. These standard deviation maps reflect not only the propagated errors from distance and peculiar velocity measurement uncertainties but also the systematic uncertainties introduced by the HMC reconstruction used for the peculiar velocity correction.

\begin{figure}[h!]
\begin{center}
\includegraphics[width=0.9\columnwidth,angle=-0]{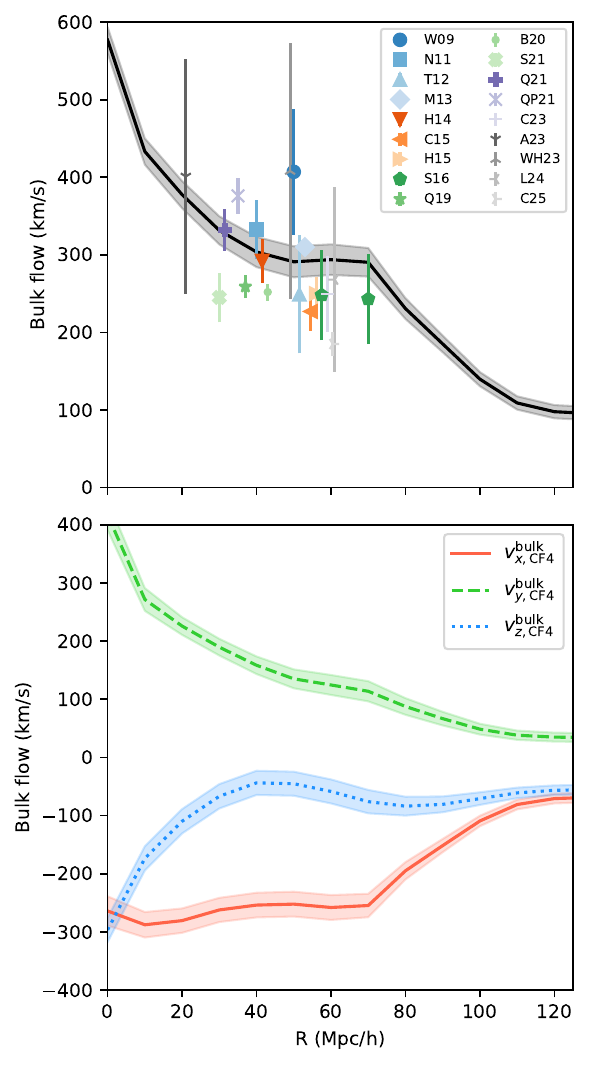}
\caption{Bulk flow of the reconstructed peculiar velocity field. Top: bulk flow of the norm of the velocity field, with the black solid line representing the mean across all 1,000 realizations and the transparent bands indicating the standard deviation. Bulk flow measurements from the literature, using supernovae or galaxy peculiar velocity datasets, are also shown. Bottom: SGX 
(red solid line), SGY 
(green dashed line), and SGZ 
(blue dotted line) components of the bulk flow, 
with the same representation for mean and standard deviation (transparent bands of the same colors).}
\label{fig:CF4bulkflow}
\end{center}
\end{figure}

The bulk flow of the reconstructed peculiar velocity field, computed with equation \ref{eq:bulkflow}, is illustrated in Figure \ref{fig:CF4bulkflow}. The top panel shows the bulk flow of the norm of the velocity field, with the black solid line representing the mean across all 1,000 realizations, while the transparent bands indicate the standard deviation over the same realizations. Bulk flow measurements from the literature, using supernovae or galaxy peculiar velocity datasets, are also added (W09: \cite{2009MNRAS.392..743W}, N11: \cite{2011ApJ...736...93N}, T12: \cite{2012MNRAS.420..447T}, M13: \cite{2013MNRAS.428.2017M}, H14: \cite{2014MNRAS.445..402H}, C15: \cite{2015MNRAS.450..317C}, H15: \cite{2015MNRAS.449.4494H}, S16: \cite{2016MNRAS.455..386S}, Q19: \cite{2019MNRAS.482.1920Q}, B20: \cite{2020MNRAS.498.2703B}, S21: \cite{2021MNRAS.505.2349S}, Q21: \cite{2021RAA....21..242Q}, QP21: \cite{2021ApJ...922...59Q}, C23: \cite{2023A&A...670L..15C}, A23: \cite{2023BrJPh..53...49A}, WH23: \cite{2023MNRAS.526.3051W}, L24: \cite{2024ApJ...967...47L}, C25: \cite{2025arXiv250201308C}). These measurements, taken from Table 1 in \cite{2025arXiv250201308C}, are in agreement with our results, further validating the accuracy of the reconstructed field. The bottom panel displays the individual components of the bulk flow along the SGX 
(red), SGY 
(green), and SGZ 
(blue) axes, 
with the mean again represented by a solid line and the standard deviation by transparent bands. The shape of the norm of the bulk flow and its components also matches the results found in \cite{2025arXiv250201308C}, confirming the consistency of the model with previous findings.

\begin{figure*}[ht!]
\begin{center}
\includegraphics[width=\textwidth,angle=-0]{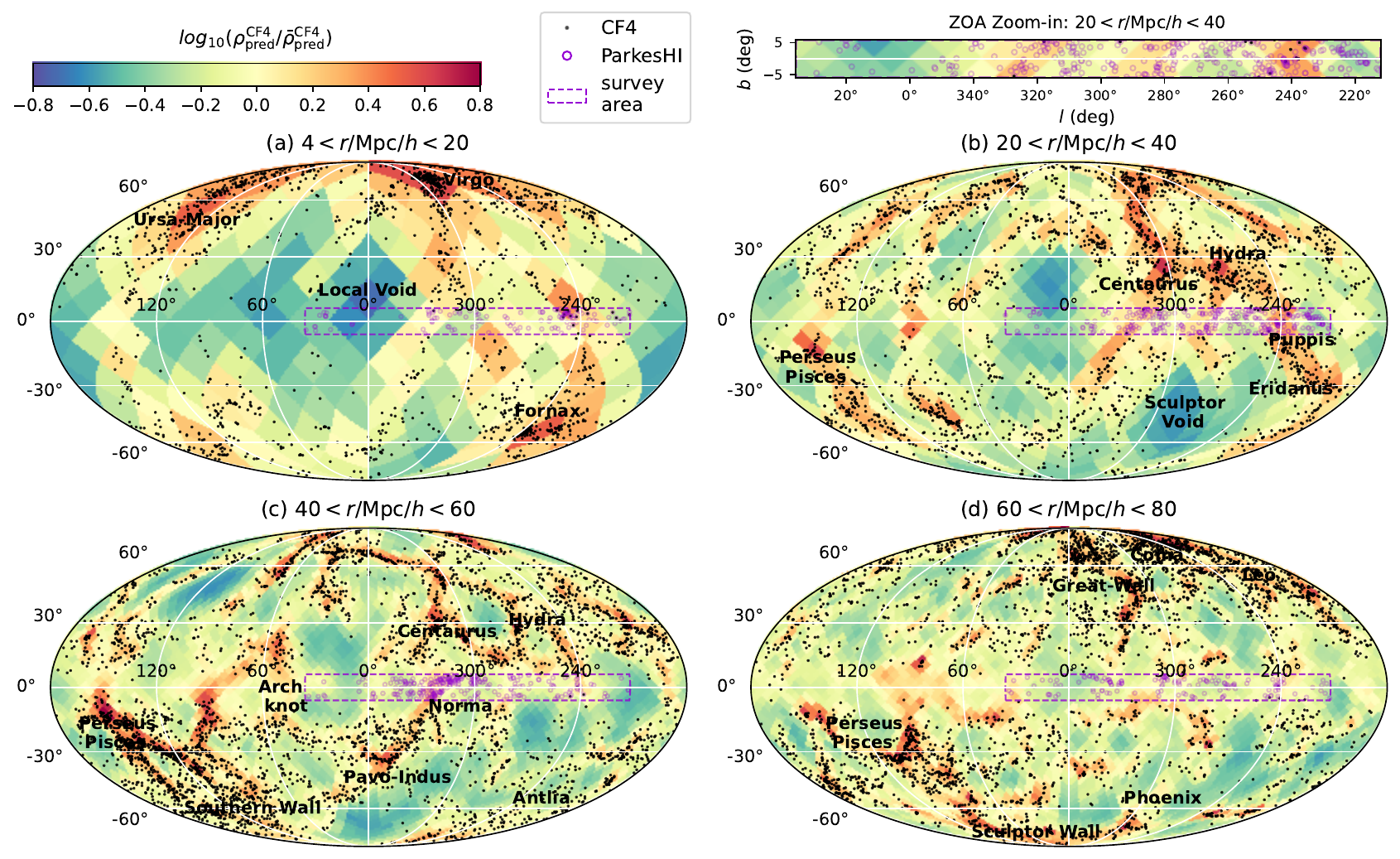}
\caption{Aitoff skymaps in Galactic coordinates showing the mean reconstructed density field across 1,000 realizations at different radial shells: (a) 4 to $20\,{\rm Mpc}/h$, (b) 20 to $40\,{\rm Mpc}/h$, (c) 40 to $60\,{\rm Mpc}/h$, and (d) 60 to $80\,{\rm Mpc}/h$. The density field in each skymap is normalized by the mean density of that shell. The color scheme ranges from blue (underdensities/voids) to red (overdensities/clusters). Tiny black dots represent CF4 galaxies. Empty purple circles indicate galaxies from the Parkes H\textsc{i} Zone of Avoidance Survey. 
A small inset at the top of panel (b) shows a zoom-in on the H\textsc{i} survey area in the ZOA within the 20--$40\,{\rm Mpc}/h$ shell, using the same color and marker scheme as the main panels.} 
\label{fig:CF4skymaps}
\end{center}
\end{figure*}

\subsection{Uncovering structures in the ZOA}

Figure \ref{fig:CF4skymaps} presents four Aitoff skymaps in Galactic coordinates, each depicting the mean reconstructed density field across all 1,000 realizations, at different radial shells of the same width. The density field in each skymap is normalized by the mean density of that shell. Panel (a) corresponds to the radial range from 4 to $20\,{\rm Mpc}/h$, panel (b) spans 20 to $40\,{\rm Mpc}/h$, panel (c) covers 40 to $60\,{\rm Mpc}/h$, and panel (d) represents 60 to $80\,{\rm Mpc}/h$. 
A small inset at the top of panel (b) provides a zoom-in view of the H\textsc{i} survey area within the ZOA in the 20–$40\,{\rm Mpc}/h$ shell. 
The color scheme follows that of Figure \ref{fig:CF4cosmography}, with blue indicating underdensities (voids) and red representing overdensities (clusters). Density features are annotated across all four shells with the names of their associated cosmic structures. Tiny black dots represent CF4 galaxies, which are in good agreement with the reconstructed overdensities, as expected, since the CF4 data was used in the reconstruction. 

The empty purple circles represent galaxies from the Parkes H\textsc{i} Zone of Avoidance Survey \citep[HIZOA, ][]{2016AJ....151...52S}, a blind 21-cm survey targeting H\textsc{i}-rich galaxies obscured by the Galactic plane. The survey covers Galactic longitudes $l \sim 36^\circ–212^\circ$ and latitudes $|b| \lesssim 6^\circ$, with 883 galaxies detected up to $cz \sim 12{,}000\ \mathrm{km/s}$. Within our analysis volume, the dataset is relatively complete in the 20–60 Mpc/$h$ range, while the 2–20 Mpc/$h$ bin contains fewer galaxies due to volume effects, and coverage declines above 60 Mpc/$h$ because of survey limitations. Similarly, we observe a decline in data coverage above 60 Mpc/$h$, due to survey limitations. 
Despite these variations, clusters of galaxies detected in this ZOA survey align with the reconstructed overdensities, despite the fact that this dataset was not used in the reconstruction and the CF4 data contains few to no galaxies in the ZOA region. 

To assess whether HIZOA galaxies preferentially trace overdense regions in our reconstruction, we measured the mean density at the galaxy positions and compared it to the mean density at randomly selected points within the same survey volume and redshift distribution. We performed this analysis in the same four radial shells as in Figure \ref{fig:CF4skymaps}. In the innermost shells, H\textsc{i} galaxies are strongly biased toward overdense regions: the mean density at their positions is significantly higher than for random points (4–20 Mpc/$h$: 2.12 vs. 0.53; 20–40 Mpc/$h$: 1.28 vs. 0.67). In the outer shells, the mean densities at HIZOA galaxy positions are consistent with or slightly below random expectations (40–60 Mpc/$h$: 0.79 vs. 0.77; 60–80 Mpc/$h$: 0.61 vs. 0.66). These results confirm that, at least in the nearby Universe, H\textsc{i} galaxies predominantly occupy overdense regions in our reconstruction.
This alignment highlights the potential of galaxy peculiar velocities as a powerful tool for probing the total matter distribution in the universe, even in regions with sparse observational data, such as the ZOA.  It also means that our reconstruction of the peculiar velocity field enables the identification and characterization of structures in these sparse regions, such as the GA located within the ZOA. 

\begin{figure*}[ht!]
\begin{center}
\includegraphics[width=0.9\textwidth,angle=-0]{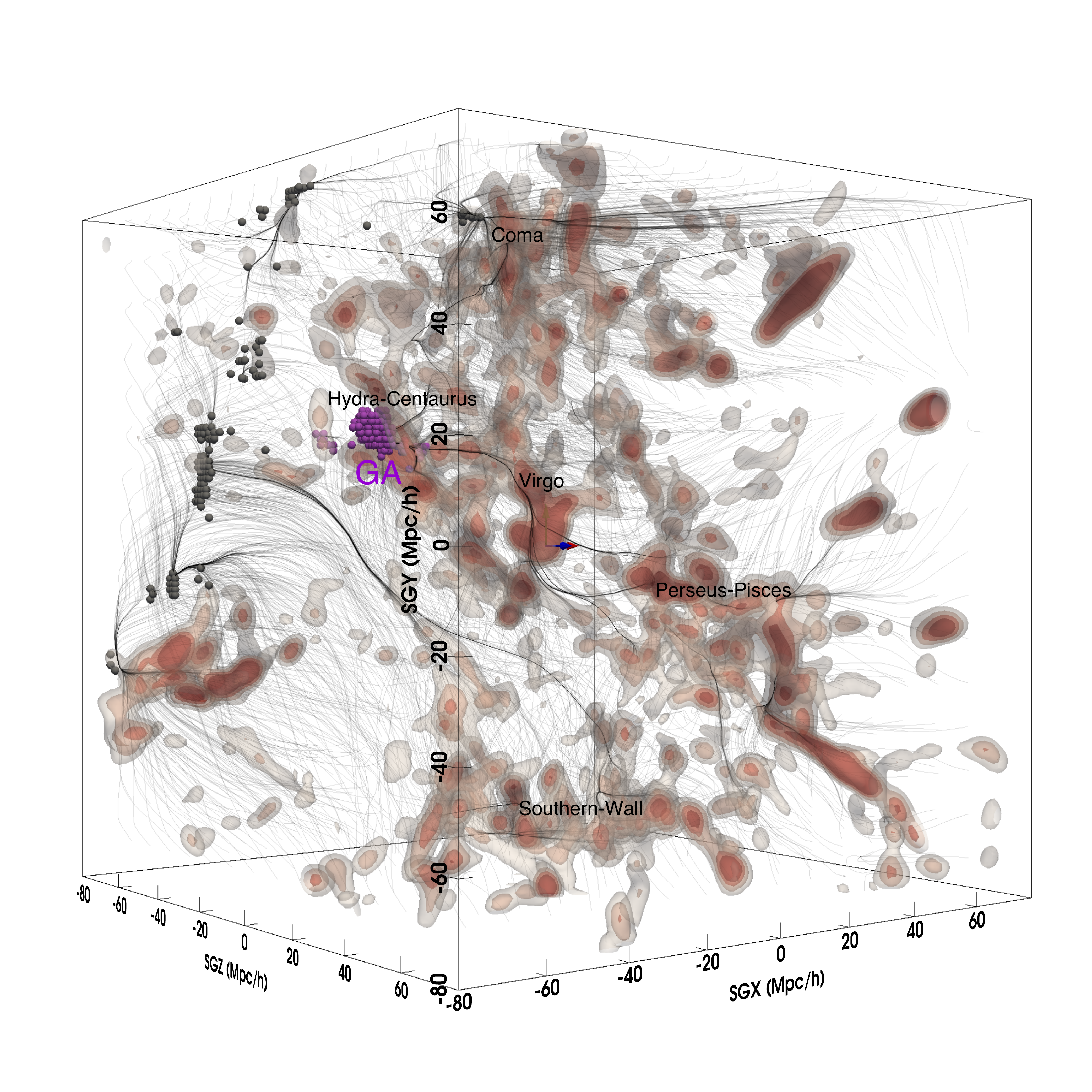}
\caption{3D visualization of the mean density and peculiar velocity fields in supergalactic coordinates. The isosurfaces represent the mean density field across 1,000 realizations, with three levels highlighting different density thresholds. Streamlines depict the mean peculiar velocity field, also averaged over 1,000 realizations. Tiny purple spheres mark the positions of GA candidates. Tiny grey spheres indicate the positions of streamline endpoints that were not identified as GA, where the streamlines terminated at the box limits. The orientation and scale of the visualization are shown by the SGX, SGY, and SGZ axes, represented by red, green, and blue arrows at the observer’s position (the cube center), each with a length of 10 Mpc/$h$. An interactive version of this visualization is available on \href{https://skfb.ly/pCUVy}{Sketchfab}.}
\label{fig:GAcandidates}
\end{center}
\end{figure*}

\subsection{Looking for the Great Attractor}

We search for the GA in our velocity field reconstruction because it is expected to be the dominant attractor influencing local galaxy flows, historically associated with the Centaurus, Hydra, and Norma clusters. Our CNN-based reconstruction provides higher resolution and captures nonlinear structures, thanks to its training on $N$-body simulations, allowing us to refine previous analyses of this region. In our framework, the GA corresponds to the central attractor of Laniakea \citep{2014Natur.513...71T,2023A&A...678A.176D}, i.e., our home supercluster and the local gravitational basin. If it is not detected at this location, this does not imply the absence of overdensities; rather, it indicates that Laniakea may be a sub-watershed within a larger basin whose center could lie elsewhere.

To identify the GA, we apply the method of \cite{2019MNRAS.489L...1D,2020MNRAS.493.3513D}, which detects attractors and repellers by analyzing streamlines obtained from the velocity field and deriving gravitational watersheds. Attractors correspond to points where streamlines converge, while repellers mark regions where flow lines diverge. The velocity field can thus be segmented into watersheds, defining basins of attraction or repulsion, where galaxies within the same basin are drawn toward a common attractor or pushed away from a repeller. The streamline originating from the Milky Way is expected to terminate at the GA location. To assess whether a GA-like structure appears in the reconstructed velocity field, we track this MW streamline and identify its associated basin of attraction and the corresponding attractor, which is then considered a possible GA candidate.

Instead of relying solely on the mean velocity field derived from the 1,000 CNN predictions, we will apply the method described below to each individual realization. This approach allows to estimate the probability of the existence of the GA and include error bars on its inferred coordinates. Prior to any further computation, a Gaussian smoothing with a radius of $1.25$ Mpc/$h$ is applied to each velocity field prediction in order to suppress non-linearities and avoid streamlines from getting stuck in local minima. Figure \ref{fig:GAcandidates} shows a three-dimensional visualization in Supergalactic coordinates of the reconstructed density field (3 levels of isosurfaces) and the reconstructed peculiar velocity field (streamlines), both averaged over all 1,000 individually smoothed realizations. The orientation and scale of the visualization are shown by the SGX, SGY, and SGZ axes, represented by red, green, and blue arrows at the observer’s position (the cube center), each with a length of 10 Mpc/$h$.

Applying the above procedure to all 1,000 realizations of the velocity field would yield at most 1,000 GA candidates, though some realizations may lack a well-defined GA, reducing the total count. The fraction of realizations in which a GA is clearly identified provides an estimate of its probability of existence. By analyzing the coordinates of all GA candidates, we determine the mean GA position along with error bars, quantifying the uncertainty in its location.

Among the 1,000 CNN velocity field realizations, a GA candidate is identified in 644 cases, corresponding to a probability of existence of 64.4\%. The GA candidates identified in these realizations are represented by purple spheres in Figure \ref{fig:GAcandidates}. The mean coordinates of these candidates are $(\mathrm{SGX}, \mathrm{SGY}, \mathrm{SGZ}) = (-42.8 \pm 4.0, 24.6 \pm 1.9, -4.0 \pm 1.7)$ Mpc/$h$, with corresponding Galactic coordinates $(l, b) = (308.4^\circ \pm 2.4^\circ, 29.0^\circ \pm 1.9^\circ)$ and a mean redshift of $cz = 4960.1 \pm 404.4$ km/s. This location aligns with the Centaurus, Hydra, and Norma clusters, in agreement with previous studies. Furthermore, our results align well with the original characterization of the GA by \cite{1988ApJ...326...19L}, which located it at $(l, b) = (307^\circ, 9^\circ)$, approximately $20^\circ$ from our estimate, and at a distance of $cz = 4350 \pm 350$ km/s, differing by about 1728 km/s.

The remaining 356 realizations exhibit streamlines that extend beyond the computational grid, suggesting a possible convergence toward structures outside the box limits, such as the Ophiuchus or Shapley Superclusters. The endpoints of these streamlines are represented by grey spheres in Figure \ref{fig:GAcandidates}. Our findings are consistent with the recent results of \cite{2024NatAs...8.1610V}.

To further validate our results, we tested an additional set of 1,000 velocity field realizations, derived from a new set of 1,000 samples of corrected CF4 peculiar velocities extracted from HMC velocity field reconstructions. The outcomes remained highly consistent, with 62.6\% of realizations identifying a GA candidate, while the remaining realizations showed streamlines converging elsewhere. This reinforces the robustness of our approach and confirms the stability of our findings across different samples. We also examined the impact of smoothing by applying a Gaussian filter with a radius of 4 Mpc/$h$. In this case, the fraction of GA candidates decreased to 43.8\%, while the remaining 56.2\% of realizations showed streamlines converging elsewhere. This behavior is consistent with the known effect of Gaussian smoothing on velocity fields, where attractors tend to fade, leading to fewer but larger basins of attraction \citep{2020MNRAS.493.3513D}.

\section{Conclusion}
\label{sec:conclusion}

In this work, we developed a CNN-based approach reconstructing the dark matter density and gravitational potential fields (and the peculiar velocity field) from bias-corrected peculiar velocities, obtained from a CF4 HMC reconstruction. Trained on $N$-body $\Lambda$CDM simulations (A-SIM), our model accurately predicts these fields, showing strong agreement with the ground truth. By applying the trained model to the CF4 data, we successfully identified known large-scale structures, even within the ZOA, and recovered a bulk flow consistent with previous studies. Since CF4 data is full of observational uncertainties and our CNN model does not currently correct for them, we use corrected peculiar velocities extracted from a CF4 reconstruction of the velocity field that accounts for observational biases. By analyzing streamline convergence in the predicted velocity field, we provided statistical evidence for the existence of the GA with a $64.4\%$ probability, locating it at Galactic coordinates $(l, b) = (308.4^\circ \pm 2.4^\circ, 29.0^\circ \pm 1.9^\circ)$ and a mean redshift of $cz = 4960.1 \pm 404.4$ km/s. This position aligns with the Centaurus, Hydra, and Norma clusters, supporting earlier findings. However, some streamlines did not converge at this location, suggesting the influence of structures beyond our analysis volume, such as Shapley or Ophiuchus.

Although our analysis relies on the CF4 peculiar velocities corrected through an existing reconstruction (HMC), the CNN model adds further value by predicting the full 3D density, potential, and velocity fields with higher resolution. Compared to the HMC-based fields, the CNN reconstruction reveals sharper filamentary structures and a more detailed cosmic web morphology, capturing nonlinear features present in the training simulations that are inaccessible to the linear HMC reconstruction. This enhanced resolution allows the CNN to resolve small-scale density and velocity variations, complementing the HMC-derived input and providing a more complete picture of the local large-scale structure.

Our approach successfully reconstructs the dark matter density and velocity fields, but it inherits systematic uncertainties from both the deep learning model and the HMC method used to derive the corrected peculiar velocities. Since our model relies on peculiar velocities from the HMC reconstruction, it inherits the systematic uncertainties of that method, which propagate through our analysis. A crucial next step is to incorporate bias correction directly into the deep learning framework, potentially during the construction of training samples, by introducing random errors that reflect the uncertainties associated with different distance measurement methodologies. This requires a deeper analysis of the CF4 dataset and mock catalogs, which we leave for a follow-up study.

Beyond addressing these limitations, several key next steps will further improve our methodology. One important goal is to extend our analysis to a larger grid, enabling a more comprehensive mapping of the entire CF4 dataset. Additionally, we aim to use our reconstructed fields as initial conditions for constrained simulations, which would allow for a more direct comparison between observations and structure formation models.

Beyond CF4, our reconstruction approach is particularly promising for upcoming peculiar velocity surveys. Unlike traditional methods that are limited by the number of available data points, our deep learning framework primarily depends on the resolution of the reconstructed grids. This adaptability makes our approach well-suited for future datasets, ensuring that as observational coverage expands, our reconstructions will continue to provide valuable insights into the large-scale structure of the universe.

\begin{acknowledgments}

Authors acknowledge Changbom Park for helpful discussion.
Authors acknowledge the Korea Institute for Advanced Study for providing computing resources (KIAS Center for Advanced Computation Linux Cluster System).
The full-sky maps in this paper have been derived using the healpy and HEALPix package.
All figures in this paper have been produced with helps of NumPy/SciPy, pandas, astropy, and Matplotlib packages.
AD is supported by a KIAS Individual Grant (PG087201) at Korea Institute for Advanced Study.
DJ is supported by NSF grants (AST-2307026,AST-2407298) at Pennsylvania State University and by KIAS Individual Grant PG088301 at Korea Institute for Advanced Study.
SEH is supported by the Korea Astronomy and Space Science Institute grant funded by the Korea government (KASA, Korea AeroSpace Administration) (2025\-186901).
HSH acknowledges the support of the National Research Foundation of Korea (NRF) grant funded by the Korea government (MSIT), NRF-2021R1A2C1094577, Samsung Electronic Co., Ltd. (Project Number IO220811-01945-01), and Hyunsong Educational \& Cultural Foundation. 
JK is supported by a KIAS Individual Grant (KG039603) via the Center for Advanced Computation at Korea Institute for Advanced Study.
HC acknowledges support from the Institut Universitaire de France and from Centre National d’Etudes Spatiales (CNES), France. 

\end{acknowledgments}

\begin{contribution}

All authors contributed to the design of the experiment, interpretation of the results, and writing of the manuscript.
AD prepared the training samples from the A-SIM simulation data; performed the deep learning training (building on the CNN architecture designed by SEH); prepared the corrected CF4 samples and applied them to the CNN; and carried out the analysis of the results.
DJ conceived the project and co-designed the analysis methodology.
SEH designed the CNN architecture and prepared the A-SIM galaxy catalog.
HSH contributed to the development of the analysis methods and interpretation of results based on CF4 and Parkes H\textsc{i} datasets.
JK co-conceived the project and designed the A-SIM simulation data, specifically the density and potential fields used as output quantities.
HC developed the HMC methodology and prepared the HMC samples used to construct the corrected CF4 datasets.

\end{contribution}

\bibliography{biblio}
\bibliographystyle{aasjournalv7}



\end{document}